\documentclass[conference]{IEEEtran}
\IEEEoverridecommandlockouts
\pdfoutput=1
\usepackage{cite}
\usepackage[hyphenbreaks]{breakurl}
\usepackage[hyphens]{url}
\usepackage{amsmath,amssymb,amsfonts}
\usepackage{algorithmic}
\usepackage{graphicx}
\usepackage{textcomp}
\usepackage{xcolor}
\usepackage[utf8]{inputenc} 
\usepackage[T1]{fontenc} 
\def\BibTeX{{\rm B\kern-.05em{\sc i\kern-.025em b}\kern-.08em
    T\kern-.1667em\lower.7ex\hbox{E}\kern-.125emX}}
\begin{document}

\title{Tubu-io Decentralized Application Development \& Test Workbench}

\author{
\IEEEauthorblockN{1\textsuperscript{st} Ercan Işık}
\IEEEauthorblockA{\textit{TUBU} \\
İstanbul, Turkey \\
ercan@tubu.io}
\and
\IEEEauthorblockN{2\textsuperscript{nd} Melih Birim}
\IEEEauthorblockA{\textit{TUBU} \\
İstanbul, Turkey \\
melih@tubu.io}
\and
\IEEEauthorblockN{3\textsuperscript{rd} Enis Karaarslan}
\IEEEauthorblockA{\textit{Department of Computer Engineering} \\
\textit{Muğla Sıtkı Koçman University}\\
Muğla, Turkey \\
enis.karaarslan@mu.edu.tr}
}
\maketitle

\begin{abstract}
Decentralized services are increasingly being developed and their proper usage in different areas is being experimented with. Autonomous codes, which are also called smart contracts, can be developed with Integrated Development Environments (IDE). However, these tools lack live environment tests. The underlying blockchain technologies are also evolving and it is not easy to catch all the developments. There is a need for an easy-to-use interface by which the developers can see the results of their codes. Tubu-io decentralized application development workbench is developed to serve as an efficient way for the programmers to deploy smart contracts on the blockchain networks and interact with them easily. It can also be used for teaching decentralized application programming for junior blockchain developers on blockchain testbeds. Finally, it will have an effect in decreasing the development time and the costs of developing decentralized application projects.
\end{abstract}


\section{Introduction}

Decentralized solutions have the potential to change how we do transactions.
Trust can be provided without an intermediary; secure and private transactions can be possible between peers. Blockchain is a decentralized ledger technology that can be used for such processes [1]. Smart contracts are introduced with Ethereum [2] that is used to form autonomous programs. These immutable codes can be called from their stored addresses and enables to run deterministic programming on the nodes of the blockchain network. Decentralized applications (DAPP) use these codes to give decentralized services. Several blockchain frameworks can be used to develop and deploy alike codes such as Ethereum [2], Hyperledger Fabric[4], Corda[5], Quorum[6] However, deploying the smart contracts on a blockchain network and interact with the system is a complex task. Tubu-io DAPP development workbench is developed to enable a user-friendly interface for the developers. 

\section{DAPP DEVELOPMENT WORKBENCH}
There are several integrated development environments (IDE) like remix, truffle and embark which provide tools in writing smart contracts[7]. However, these tools lack the interaction with the blockchain network. Tubu-io blockchain development workbench is a web platform by which the developers can easily interact with the defined blockchain networks. The developers can deploy their smart contracts to their blockchain networks and interact with them. 

The proposed system is shown in Figure 1. Tubu-io can be installed on a separate web server or one of the blockchain nodes. A separate web server would be a better choice to form a more secure design. There are two main users of the system:
\begin{itemize}
\item Blockchain Developers: The developers will use the web interface to deploy and test decentralized applications.
\item Dev Community: The developers of the Tubu-io. The project welcomes the community in developing the system.
\end{itemize}

\begin{figure}[tpb]
\begin{center}
\includegraphics[width=8cm]{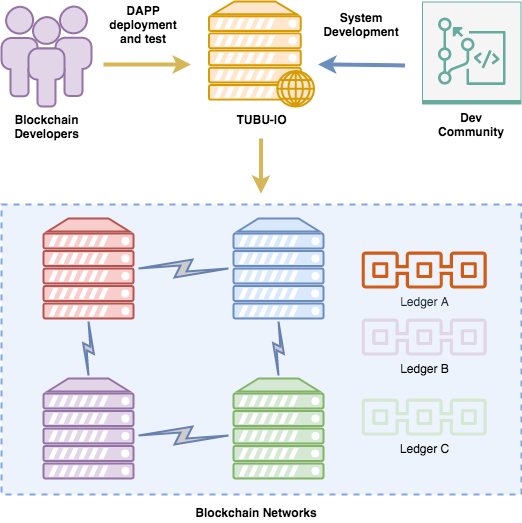}
\end{center}
\caption{Tubu-io System}
\label{fig:gui1}
\end{figure}

Tubu-io system arcitecture is shown in Figure 2. The system forms of four sub modules:
\begin{itemize}
\item Secure User/Wallet Management
\item REST API
\item Smart Contract Management
\item Version Management
\end{itemize}

\begin{figure}[tpb]
\begin{center}
\includegraphics[width=8cm]{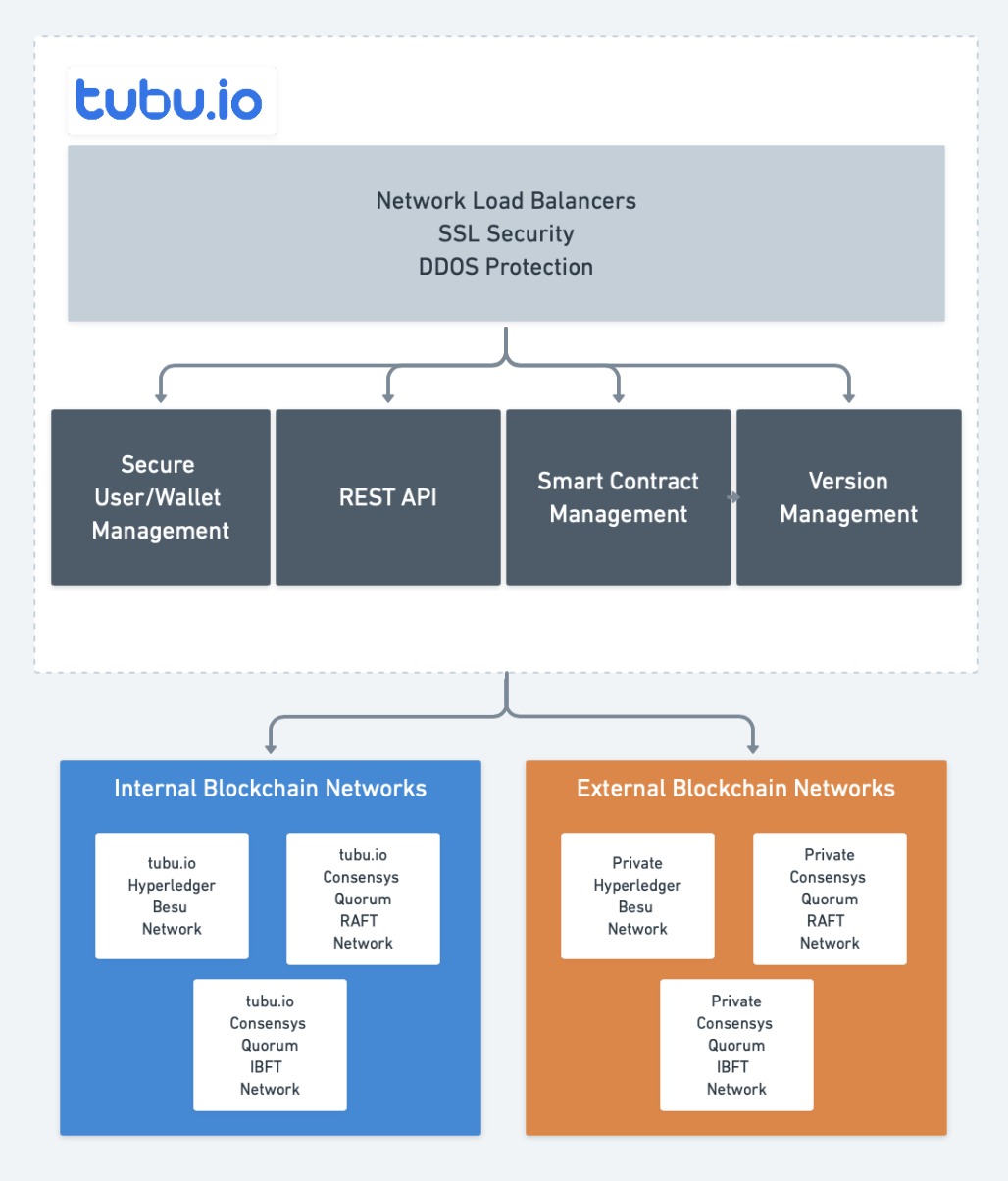}
\end{center}
\caption{Tubu-io System Architecture}
\label{fig:gui1}
\end{figure}

Tubu-io can connect to internal (defined) networks, and also external networks as well. It currently supports the Ethereum based Quorum blockchain platforms, but other blockchain platforms will be integrated in future releases. Blockchain networks are defined to the system from the user interface by adding the new network actions. Since blockchain networks need specific TCP ports to listen to the network connections, it is required to open these ports to external networks on the system and the firewall. When the available networks are defined on the system, creating the decentralized application is simplified with its web interface. The following steps should be taken:
\begin{itemize}
\item Connect to an available network
\item Create a new application
\item Deploy Contract
\item Access the deployed smart contracts via REST-API and tubu.IO SDK's
\end{itemize}

Once smart contracts are deployed, users can interact with that smart contract by calling the "Invoke” or “Call” methods. The application details section of the system provides the necessary information for this. The users can reach their blockchain network with the software development kit (SDK) of the used blockchain framework without any need for any other dependency. The current functions are the following:
\begin{itemize}
\item Share the Application: This is used to share the application with the specified system users.
\item Create API Key: API Key is required for the DAPP to interact with the contracts.
\item Deploying New Version: A new version of a pre-deployed contract can be deployed on the blockchain.
\item Accessing Contract Details: The methods, versions, accounts of the specific contract can be accessed.
\item Using SDKs: Tubu-io can be used with the SDKs that are written in widely-used programming languages such as NodeJS, Java, Python and Go.
\end{itemize}

\section{IMPLEMENTATION}
Tubu-io is developed using Javascript programming language, node.js, Vue.js, and web3.js library is used. Web3.js is a collection of Javascript API libraries developed by the ChainSafe Organization. It is used for interacting with remote Ethereum nodes. Solidity is used for the smart contracts. 

Tubu-io API is written in Javascript and allows the developers to use get and set requests, smart contract creation endpoints. The developers may use any programming language and should send requests to the Tubu-io API. Request libraries for each programming language are needed to use the Tubu-io API. SDK is developed to ease these processes for the widely used programming languages such as Python, NodeJS, Java, and Go. 

Tubu-io is installed on a web server and developers can register and then log in to the system through the web interface. The official web service can be reached from \url{https://app.tubu.io}. Projects may install Tubu-io on their private servers as well. Tubu-io web interface is shown in Fig. 2 and Fig. 3. It is used in several projects and the project repo can be reached from [10]. A detailed usage document can also be reached from [8].

\begin{figure}[tpb]
\begin{center}
\includegraphics[width=8cm]{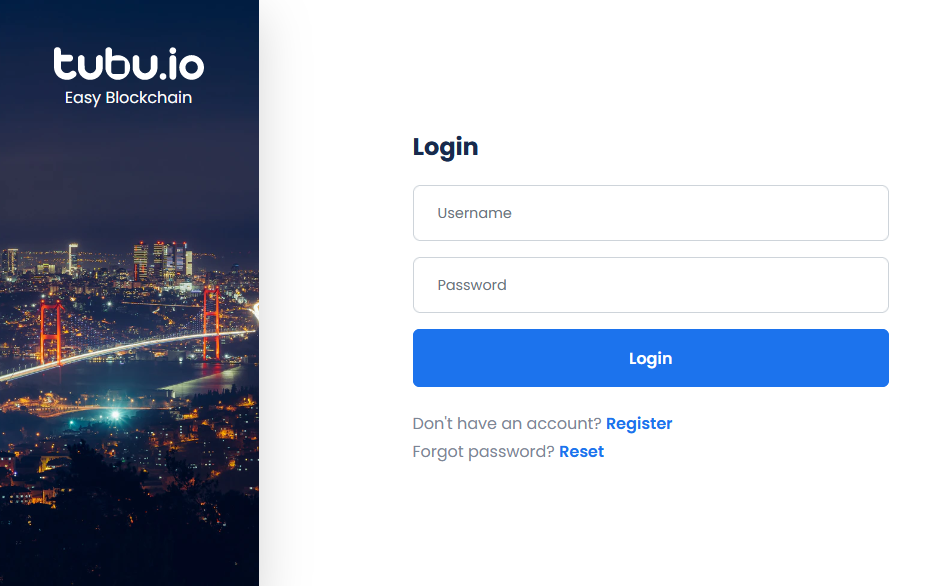}
\end{center}
\caption{Tubu-io Web Interface}
\label{fig:gui1}
\end{figure}

\begin{figure}[tpb]
\begin{center}
\includegraphics[width=8cm]{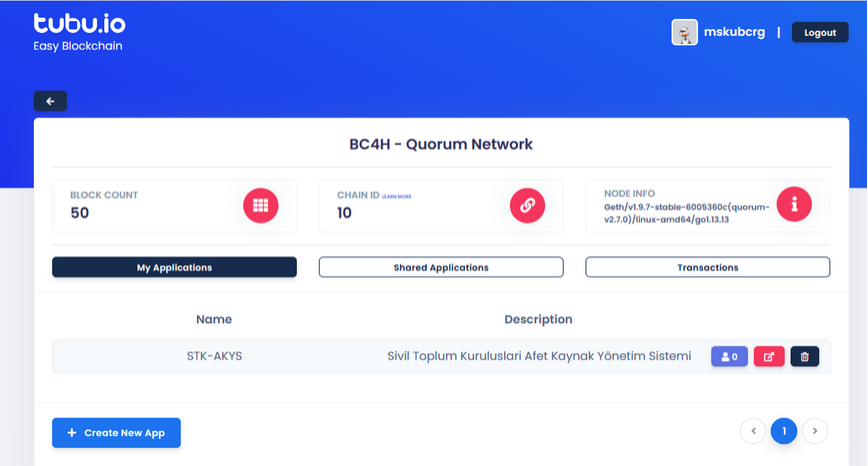}
\end{center}
\caption{Tubu-io Web Interface}
\label{fig:tubu.io.networkpage}
\end{figure}

\section{CONCLUSION}
There is a need to make blockchain solution development easier and make this process cheaper. Some countries like China see the potential and the need for this. China launched a national blockchain network in 100 cities to decrease the costs of such enterprise blockchain solutions[9].

Tubu-io decentralized application development workbench is the first step at decreasing the costs of developing blockchain projects. This tool can help non-novice users to develop decentralized solutions easier. It has been used in several projects so far. The source codes will be shared with the Apache license. It currently supports the Quorum network. Hyperledger Besu [11] is being tested and released soon. Other platforms like Hyperledger Fabric are planned to be integrated in the future. The details of the system design will be given in an upcoming paper.

\section*{ACKNOWLEDGMENT}
We would like to thank Şafak Öksüzer, Berkay, Umutcan Korkmaz, Hüseyin Emre Arı and Yunus Emre Turğut for their contributions to the development of TUBU-IO. We would also like to thank Cemal Dak for his contribution in the graphics.

\vspace{12pt}
\color{red}

\end{document}